\documentclass[a4paper]{jpconf}
\usepackage{graphicx}
\usepackage{amssymb}
\usepackage{amsmath}
\usepackage{dcolumn}
\usepackage{bm}
\usepackage{type1cm}

\begin{document}

\title{Alpha-Particle condensation  
in nuclear systems: present status and perspectives}
\author{P Schuck$^1$ $^2$ $^3$
} 
\address{$^1$Institut de Physique Nucl\'eaire, CNRS, UMR8608, \\ 
Orsay, F-91406, France\\
$^2$Universit\'e Paris-Sud, Orsay, F-91505, France \\
$^3$Laboratoire de Physique et de Mod\'elisation des Milieux Condens\'es, CNRS et, Universit\'e Joseph Fourier, 25 Av. des Martyrs, BP 166, F-38042 Grenoble Cedex 9, France
}



\begin{abstract}
$\alpha$ clustering and $\alpha$ condensation in lighter nuclei is presently 
strongly and increasingly discussed in the literature both from the 
experimental side as from the theoretical one. A discussion of the present 
status of the theory as well as outlooks for future developements will be 
presented.



\end{abstract}
\section{ Introduction} 

Since about ten years when the idea of a possible 
existence of $\alpha$ condensate type of states in $n\alpha$ nuclei  was advanced and formulated 
for the first time \cite{thsr}, many exciting new results, theoretical and experimental 
ones, have been produced. In this contribution, we would like to assess where 
we stand and what possible future extensions may be.

Let us start with the reminder that nuclear clustering and in particular $\alpha$ clustering would not exist, if we did not have in nuclear physics {\it four} different types of fermions (proton/neutron ~spin up/down), all attracting one another. We should be aware of the fact that this is a rather singular situation in fermionic many body systems. However, the possibility of future trapping of four different kinds of cold fermionic atoms, may open a new field of cluster physics with similar features as in nuclei. In a mean field description of an isolated $\alpha$ particle (what with, e.g., Skyrme forces gives reasonable results, if the c.o.m. motion is treated correctly) the four fermions can occupy the lowest 0S-level. Would there be only neutrons, only two of them can be in the 0S-level, the other two neutrons would have to be in the energetically very penalising 0P-state. That is why $\alpha$ particles exist, tetra-neutrons not. The ensuing fact is that $\alpha$ particles are very strongly bound ($E/A \sim$ 7 MeV), almost as strong as the most bound nucleus which is $^{56}$Fe ($E/A \sim$ 8 MeV). In addition the first excited state of the $\alpha$ particle ($\sim$ 20 MeV) is by factors higher than that of any other nucleus. The $\alpha$ particle can, therefore, be considered as an almost inert ideal bosonic particle. As we will see in the discussion below, inspite of its strong binding, $\alpha$ particle condensation can only exist in the so-called BEC (Bose-Einstein Condensation) phase what implies low density. There is no analogue to the BCS phase of pairing where the Cooper pairs can have very large extensions, strongly overlapping with one another, still being fully antisymmetrised. This is the reason why $\alpha$ condensation only can be present at low densities where the $\alpha$ particles do not overlap strongly (this holds, if the system consists of protons and neutrons and $\alpha$'s.
If other clusters as t, $^{3}$He,d are around, the situation may change, see below). These considerations apply for nuclear matter as well as for finite nuclei. The Hoyle state in $^{12}$C which can to a good approximation be described as a product of three $\alpha$ particles occupying all the lowest 0S state of their bosonic mean field has a density which is by a factor 3-4 lower than the one of the ground state of $^{12}$C. In the ground state there exist $\alpha$-type of correlations but there is no condensation phenomenon. Let us start our considerations with infinite matter.

\section {$\alpha$ particle condensation in infinite matter.} 
The in medium four-body equation can be written in the following form

\begin{equation}
(E_{\alpha,{\bf K}} - \varepsilon_1 - \varepsilon_2-\varepsilon_3-\varepsilon_4)
\Psi^{\alpha,{\bf K}}_{1234}=(1-f_1-f_2)v_{121'2'}\Psi^{\alpha,{\bf K}}_{1'2'34}
                        +(1-f_1-f_3)v_{131'3'}\Psi^{\alpha,{\bf K}}_{1'23'4}
                        + ...
\end{equation}

In total, there are six terms coming from  permutations. The $\varepsilon_i$ are kinetic energies plus mean field corrections; $v_{1234}$ are the matrixelements of the two body interaction, and $f_i$ is a Fermi-Dirac distribution of the uncorrelated nucleons accounting for phase space blocking. Repeated indices are summed over and index numbers comprise momenta and spins. The above equation considers, therefore, {\it one} quartet in a gas of uncorrelated nucleons at temperature $T$. The analogous two body equation can be used to determine the critical temperature $T_c$ for the onset of superfluidity or supraconductivity where $T_c$ has to be determined so that the eigenvalue comes at two times the chemical potential $\mu$. This is the famous Thouless criterion of BCS theory. In analogy with pairing, one has to find the critical temperature $T_c^{\alpha}$ so that the eigenvalue of the four body equation (1) comes at $4\mu$. The in medium four body equation is very difficult to solve. Nontheless, the solution has been found employing the Faddeev-Yakubovsky equations and using the Malfliet-Tjohn bare nucleon-nucleon interaction which yields realistic nucleon-nucleon  phase shifts and properties of an isolated $\alpha$ particle \cite{Sogo1}.
To simplify the problem, we made in addition a very easy to handle variational ansatz of the four body wave function in (1). It consists of a mean field ansatz for the $\alpha$ particle projected on good total momentum. In momentum space this is

\begin{equation}
\Psi_{1234} \propto \delta({\bf K} - {\bf k}_1- {\bf k}_2- {\bf k}_3- {\bf k}_4)
\varphi({\bf k}_1)\varphi({\bf k}_2)\varphi({\bf k}_3)\varphi({\bf k}_4)
\end{equation}

Inserting this ansatz into (1), one obtains a nonlinear HF-type of equation for 
the S-wave function $\varphi({\bf k})$. Of course, for quartet condensation, we choose ${\bf K}=0$. With the mean field ansatz (2), one cannot use a bare force. We adjusted an effective separable force with two parameters which are chosen to reproduce binding energy and radius of the free $\alpha$ particle. The full Faddeev-Yakubovsky solution of (1) is shown for symmetric and asymmetric matter in Fig. 1 (crosses).
\begin{figure}[h]
\hspace{0.7cm}
\begin{minipage}[h]{7.cm}
\includegraphics[width=6.5cm]{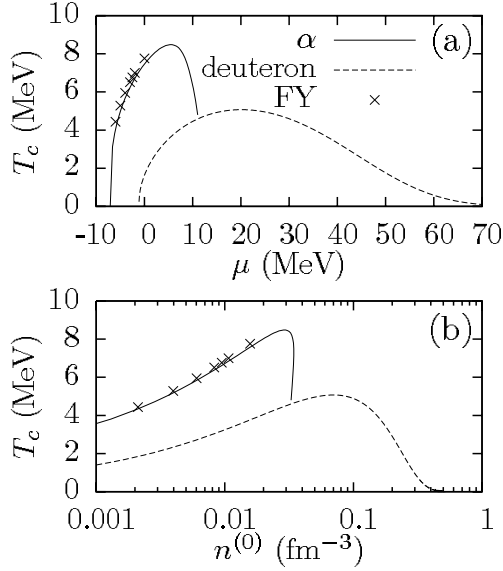}
\end{minipage}
\hfill
\hspace{-0.5cm}
\begin{minipage}[h]{6.5cm}
\caption{Critical temperatures for $\alpha$ particle and deuteron condensation 
in symmetric nuclear matter as a function of $\mu$ (a) and 
density $n^{(0)}$ (b).}
\label{Tc-sym}
\end{minipage}
\end{figure}
We see that the ansatz (2) which very much eases the otherwise difficult solution of (1) works very well (continuous line). Also shown is the critical temperature for deuteron condensation. The striking feature is that $\alpha$ particle condensation abruptly breaks down already at very low density which approximately coincides with the point where the $\alpha$'s start to overlap appreciately (this fact was already found in \cite{RSSN} using a somewhat different variational ansatz for $\Psi_{1234}$). On the other hand deuteron condensation goes on up to very high densities and the limit is only triggered by the range of the effective force (which has been readjusted to reproduce the deuteron properties). This is so for symmetric nuclear matter. For strong asymmetry, deuteron condensation breaks down earlier than $\alpha$ condensation because the $\alpha$ particle according to its much stronger binding is less sensitive to asymmetry \cite{Sogo2}. This is the  afore mentioned phenomenon that $\alpha$ condensation only exists in the BEC phase, i.e., at low density, whereas deuteron continuously goes from negative to positive chemical potentials where for the latter the deuterons turn into large size Cooper pairs. More on this can be found in \cite{Sogo1} \cite{RSSN}. We should mention that our calculation of $T_c^{\alpha}$ is only reliable rather close to the break down point. For lower densities, the $T_c^{\alpha}$ should join the one for condensation of ideal bosons ($\alpha$'s). To describe this feature, one should extend our theory to the so-called Nozi\`eres Schmitt-Rink (NSR) theory \cite{NSR} for pairing, see also \cite{urban}, to $\alpha$ particle condensation.
This, however, has not been worked out so far and remains a task for the future.

At zero temperature, there are many $\alpha$'s which go into the condensate 
phase. For this, we have to set up an approach analogous to the nonlinear BCS 
theory. Equation (1) corresponds to the linearised version and only 
describes {\it one} $\alpha$ particle in an otherwise uncorrelated gas (at 
finite $T$) of fermions. In finite nuclei, there may exist such a situation 
even at zero temperature. This is the case of $^{212}$Po which can to a certain 
extent be viewed as an $\alpha$ particle sitting on top of the doubly 
magic core 
of $^{208}$Pb which can be well described by a HF-mean field approach, i.e. a 
Fermi gas in a container. We will come back to this later when we discuss 
finite nuclei.

After having considered the linearised version of the equation for the quartet order parameter at the critical temperature, let us now try to write down, in analogy to the BCS case, the fully non-linear system of equations for the quartet order parameter. To clearly see the analogy to the BCS case, let us repeat the latter equations in a slightly unusual way. The pairing order parameter allowing for non-zero c.o.m. momentum of the pairs, reads

\begin{equation}
(\varepsilon_{k_1} + \varepsilon_{k_2})\kappa_{{\bf k}_1{\bf k}_2} + (1-n_{k_1}-n_{k_2})
\Delta_{{\bf k}_1{\bf k}_2} = 2\mu \kappa_{{\bf k}_1{\bf k}_2}
\end{equation}

where $\Delta_{{\bf k}_1{\bf k}_2} = \sum 
v_{k_1k_2k_{1'}k_{2'}} \kappa_{{\bf k}_{1'}{\bf k}_{2'}}$ and   $\kappa_{{\bf k}_1{\bf k}_2} = \langle \mbox{BCS}|c_{{\bf k}_1}c_{{\bf k}_2}|\mbox{BCS}\rangle =
u_{{\bf k}_1}v_{{\bf k}_2}$ is the usual pairing tensor (with spin indices 
suppressed) and $n_k=v_k^2 = 1-u_k^2$ are the BCS occupation numbers.
The $\varepsilon_k$'s are, as before, the kinetic energies, eventually including a 
mean field correction. The occupation numbers can be obtained from the Dyson 
equation

\begin{equation}
G_k^{\omega} = G_k^0 + G_k^0 M_k^{\omega}G_{k}^{\omega} ~~~~\mbox{with}~~~~~
M_{k} = \frac{\Delta_k\Delta^*_k}
{\omega + \varepsilon_{ k}}
\end{equation}

the BCS mass operator where $\Delta_k$ is the diagonal part of the gap for cases where the pairs are at rest. From the single particle Green's function, obviously we can calculate the occupation numbers closing, thus, the typical BCS selfconsistency cycle. Inspired by the BCS case, we then write for the quartet order 
parameter, see (1)

\begin{equation}
(4\mu - \varepsilon_1 - \varepsilon_2 -\varepsilon_3 -\varepsilon_4)
\kappa_{1234} = (1-n_1-n_2)v_{121'2'}\kappa_{1'2'34}
             + (1-n_1-n_3)v_{131'3'}\kappa_{1'23'4}
             + ...
\end{equation}

with $\kappa_{1234} = \langle c_1c_2c_3c_4\rangle$.
Again, we have to close the equation with the Dyson equation for the occupation numbers. However, the massoperator now contains the quartet order parameter
\begin{equation}
M_1^{\alpha}=\sum_{234}\frac{\Delta_{1234} [\bar n_2^0 \bar n_3^0
\bar n_4^0 + n_2^0n_3^0n_4^0] \Delta^*_{1234}}{\omega +\varepsilon_2+\varepsilon_3+\varepsilon_4}\delta({\bf k}_1 + {\bf k}_2 + {\bf k}_3 + {\bf k}_4)
\end{equation}

with $\bar n^0 = 1 - n^0$ and $n_i^0$ being the uncorrelated Fermi function, i.e., the Fermi step and $\Delta_{1234}= \sum V_{123'4'}\kappa_{3'4'34}$ 
where $V_{1234}$ is an effective coupling vertex linking the single particle 
motion to the order parameter (for more details and derivation, see \cite{Sogo3}). Before trying to solve this equation, let us discuss the differences between the pairing 
and the quartet case. 
The first thing which strikes is that the three 'holes' only have to have total momentum ${\bf k}_2+{\bf k}_3+{\bf k}_4 = -{\bf k}_1$ and, therefore, we have a remaining sum over momenta. In the pairing case with only one 'hole', there is no sum. Furthermore in the pairing case the hole propagator has no phase space factor because 'forward' and 'backward' going parts add up to one: $\bar n^0 + n^0 =1$. In the quartet case there are three hole propagators and the corresponding sum of phase space factors does NOT add up to one, i.e. 
$\bar n^0_1\bar n^0_2\bar n^0_3 +n^0_1n^0_2n^0_3 \ne 1$!! This makes a dramatic difference with the pairing case. In order to understand this a little better, let us compare the level density of a single hole with the one of three holes:
\begin{equation}
g^{1h}(\omega )=\sum_k[\bar n^0_k + n^0_k]\delta(\omega+\varepsilon_k)
=\sum_k \delta(\omega+\varepsilon_k)
\end{equation}

\begin{equation}
g^{3h}(\omega )=\sum_{{\bf k}_1{\bf k}_2{\bf k}_3}[\bar n^0_{k_1}\bar n^0_{k_2}\bar 
n^0_{k_3} + n^0_{k_1} n^0_{k_2} n^0_{k_3}]\delta(\omega+\varepsilon_{k_1}+
\varepsilon_{k_2}+\varepsilon_{k_3})
\end{equation}

The 3h level density is shown in Fig.2. 
\begin{figure}[h]
\parbox[t]{5cm}{\vspace{0cm}
\includegraphics[width=5.cm]{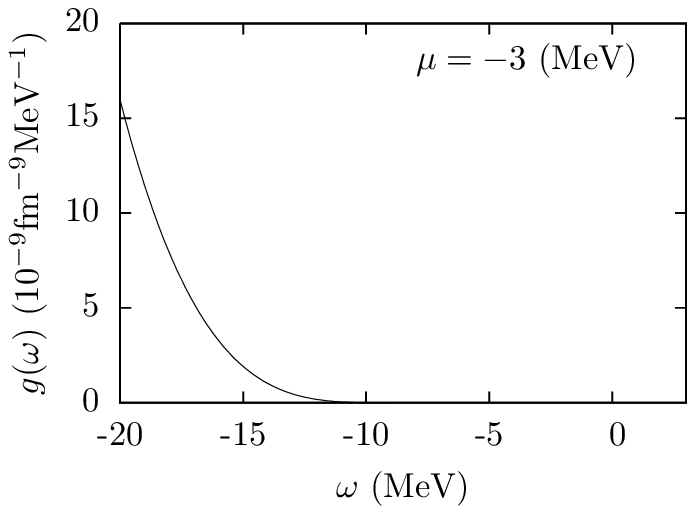}
}
\hspace{3mm}
\parbox[t]{5cm}{\vspace{0cm}
\includegraphics[width=5.cm]{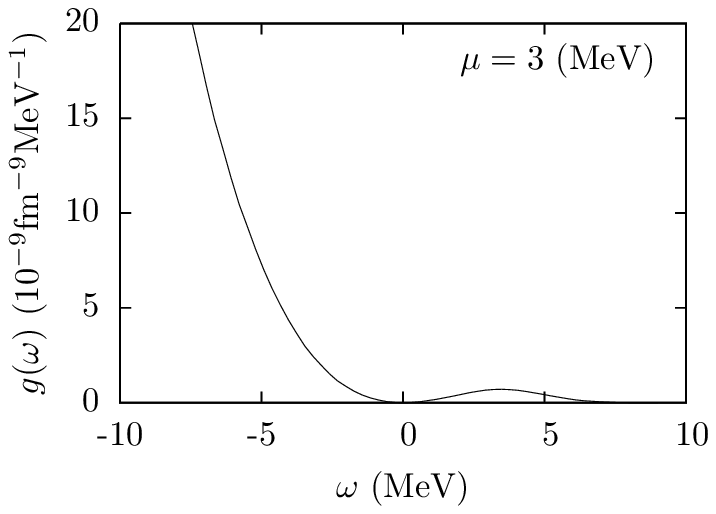}
}
\hspace{3mm}
\parbox[t]{4.7cm}{\vspace{0cm}
\caption{3h level densities for negative and positive chemical potentials, respectively. Note that on the horizontal axis, the origine corresponds to $\mu=0$. }
\label{fig:4}}
\end{figure}
We see that for positive $\mu$ there is a striking difference with the 1h level density. At positive $\mu$, $g^{1h}(\omega = \mu)$ is obviously finite (not shown), whereas $g^{3h}(\omega = 3\mu)$ goes through zero. This is because phase space constraints and energy conservation cannot be fullfilled simultaneously at the Fermi surface in the latter case as is easily verified. That is, in the quartet case, exactly at the point where the correlations should be built up, namely at the Fermi-level, there is no level density! As a consequence, no quartet condensation is possible for positive $\mu$. On the contrary for negative $\mu$, $n^0=0$ and, thus, the phase space factor in the case of $g^{3h}$ is also equal to one and then there is no qualitative difference with the $1h$ case. This explains in a natural way why quartet condensation is not possible at positive chemical potential.
It is by the way well known that any $mp-nh$ level density, besides the single particle case, goes through zero at the Fermi level. For example, the origin of fermions at the Fermi energy having an infinite free mean path stems from the fact that the 2p-1h (2h-1p) level density is zero at the Fermi level what is the equivalent to the imaginary part of the optical potential being zero at that energy. Also the 2p-2h level density which plays an important role for the damping of zero sound modes is zero at zero frequency. As first pointed out by Landau, it approaches zero as $\omega ^2$. In conclusion for positive 
$\mu$ only pairing survives whereas quartetting breaks down and only 
exists in the BEC phase with negative $\mu$. The situation may be different when other light clusters are present, i.e. in a mixed gas of, e.g., nucleons, tritons ($^3$He), deuterons. Then a nucleon with momentum ${\bf k}$ may, eventually,  directly pair with, e.g., a triton of momentum $-{\bf k}$ (or the other way round), rather similar to the standard pairing situation besides the fact that now two fermions with different masses have to pair up. Similar considerations hold for the pairing of two deuterons (pairing of 'bosons'). In compact star physics such situations may exist when the star is cooling down. The extension of our theory to this scenario is a task for the future. The full solution of the nonlinear set of equations (6) and (7) is again very much eased in taking for the order parameter the factorisation ansatz (2). The most interesting result is that the occupation numbers,e.g., for $\mu$ around zero is far from being close to saturation. At $k=0$ it is only approximately $n_{k=0} \sim 0.30$. This scenario is analogous to pairing in the BEC regime. More results can be found in \cite{Sogo3}.

\section{Finite Nuclei} 
As we know from pairing, a direct observation of 
condensation phenomena only is possible in finite nuclei. Of course, in such 
small systems, there cannot exist a condensation in the macroscopic sense. 
Nevertheless, as we know very well, only a handfull of Cooper pairs suffices 
to show clear signatures of pairing in nuclei. For $\alpha$ particle 
condensation it is the same story. We also only can expect that there exists 
about a handfull of $\alpha$ particles, essentially in lighter $n\alpha$ 
nuclei, in a gaseous phase at low density. It is, indeed, surprising that 
such states at low density with about $\rho = \rho_0/3-\rho_0/4$ with $\rho_0$ 
the density at saturation, indeed exist as excited quite long lived states in 
those nuclei. The most fameous example is the Hoyle state in $^{12}$C 
at 7.65 MeV, just about 300 keV above the 3$\alpha$ threshold.
We will not dwell much on the successful theoretical description of this state 
(and others,e.g., in $^{16}$O) 
with the THSR wave function \cite{thsr}, since this has been presented in the 
recent literature a great number of times \cite{book}. Let us only make a couple of 
remarks. The THSR wave function is schematically written for a finite number of quartets as 

\begin{equation}
\Psi^{\mbox{THSR}}_{n\alpha} = {\mathcal A}[\Phi_{\alpha}\Phi_{\alpha}....\Phi_{\alpha}]
\end{equation}

where the single $\alpha$ wave function $\Phi_{\alpha}$ depends on four spatial 
coordinates, the spin-isospin part being suppressed. This wave function is 
fully antisymmetric due to the antisymmetriser $\mathcal A$ and is analogous 
to the number projected BCS wave function

\begin{equation}
\Psi^{\mbox{BCS}} = {\mathcal A}[\phi_{\mbox{pair}}\phi_{\mbox{pair}}.....\phi_{\mbox{pair}}]
\end{equation}

where $\phi_{\mbox{pair}}$ is the Cooper pair wave function depending on two 
spatial 
coordinates. The calculus with 
the $\alpha$ condensate wave function is very much facilitated with a 
variational ansatz where $\Phi_{\alpha}$ is split into a product of a c.o.m. 
gaussian with a large width parameter $B$ times another, intrinsic, gaussian 
depending only
on the relative coordinates of the $\alpha$ particle and having a width 
parameter $b$ which corresponds to the size of an isolated $\alpha$ particle. 
The first important remark to be made is that this THSR wave function contains 
two important limits: if $B=b$ then it corresponds to a pure harmonic 
oscillator Slater determinant. If $B >> b$, then the $\alpha$'s are so distant 
from one another that the Pauli principle among the different $\alpha$'s can be 
neglected and, thus, the antisymmetriser be dropped. The THSR wave function is 
then a pure product state of $\alpha$ particles, i.e., a condensate state of 
ideal bosons. Reality is, of course, in between those limits and one important 
task is to find out whether reality is closer to a Slater determinant or to a 
Bose condensate. That this question must be carefully investigated, can also 
be deduced from the fact that a number projected BCS wave function always 
leads to a non trivial pairing solution. For example for $^{208}$Pb, one obtains a non 
trivial BCS solution inspite of the fact that $^{208}$Pb is certainly not 
superfluid. Only when the original particle number breaking BCS theory
with $|\mbox{BCS}\rangle = e^{\sum z_{kk'}c_k^+c_{-k}^+}|vac\rangle$ has a non-trivial 
solution, we can speek of a superfluid nucleus. For $^{208}$Pb there is no such 
solution. One way to analyse whether the THSR approach leads primarily to 
an $\alpha$ condensate or to a Slater determinant, is to investigate the 
bosonic occupation numbers. An ideal Fermi gas has occupation numbers which 
are one or zero. In an ideal Bose condensate the bosons will occupy the lowest 
single particle state with 100 $\%$. Of course, in real nuclei, neither the 
fermionic occupation numbers nor the bosonic occupations are the ones of an 
ideal gas. For nucleons, we know that occupation numbers are around 70-80\% 
down from unity. For the $\alpha$ occupations in the condensate, we also 
have found  a number around 70\%, all the other 
occupancies being down by at least a factor of ten, whereas 
for the ground state the occupation numbers are almost equally distributed 
according to the SU3 shell model scheme \cite{book}  . This is a typical 
condensate situation, though not totally the one of an ideal condensate. Residual antisymmetry 
effects between the slightly overlapping $\alpha$ particles makes that 
with a 
certain probability the $\alpha$'s are scattered out of the condensate. In 
addition there are also other correlations at work. One of the most 
important one is certainly the formation of $^8$Be clusters out of two 
$\alpha$'s. It is even not clear whether an $\alpha$ gas does not in reality 
consist out of a gas of $^8$Be's. The latter are, of course also bosons and 
the old question arises whether in an attractive Bose gas the bosons 
condensate as singles or as molecules \cite{noz-jam}. This is certainly a very interesting 
question which desserves further studies in the future.

Many other extensions of $\alpha$ condensation are presently discussed 
theoretically and experimentally. An intersting aspect is whether on top of 
the $\alpha$ condensate states excited $\alpha$ gas states exist. A long 
debate has recently been closed about the nature and existence of the 
second $2^+$ state close to 10 MeV excitation energy in $^{12}$C. Very nice 
experimental results by M. Freer, M. Itoh, and M. Gai have recently shown that 
this state is there and that it is part of a family of $\alpha$ gas 
states \cite{Gai}\cite{itoh}. Many more results, for instance, in $^{16}$O are to be 
expected. One of the most exciting aspects is that one may be able to 
dismantle $n\alpha$ nuclei with $n$ a rather large number like $n=10$ or more 
into $n$ $\alpha$ particles. First results in this direction have been 
reported at this conference with $^{56}$Ni, i.e. a nucleus with 14 $\alpha$ 
particles by H. Akimune. A dream would be that all $\alpha$ particles be just 
excited 
to the Ikeda threshold and then they  disintegrate in a very slow 
motion as a kind of coherent state driven by the Coulomb force, i.e., it would 
be some kind of soft Coulomb explosion. This would then be rather close to what 
happens with a trapped Bose condensate of cold atoms upon switching off  the 
trapping potential. Other exciting perspectives are that $\alpha$ particles 
could exist in a gaseous phase on top of an inert core. For example four 
$\alpha$'s on top of $^{16}$O in $^{32}$S, or other variants, even in quite 
heavy nuclei. Because of space restrictions, we, unfortunately, cannot extend 
further on these exciting aspects of $\alpha$ gas type of states in nuclei.

 Before closing, we would like to discuss, however, the question of a possible
preformation of $\alpha$ particles in heavy nuclei. As well known, this 
question is of great importance fo the description of $\alpha$ decay rates. 
Let us take the example already alluded to in the infinite matter section, namely $^{212}$Po. Since the 
lead core is doubly magic,
one can treat it in mean field, i.e. as a Slater determinant with a suitable 
Skyrme or Gogny type of force. That is we may view $^{212}$Po as an $\alpha$ particle on top of a finite Fermi gas. If such a configuration exists, and the $\alpha$ decay rate of $^{212}$Po suggests that, it is clear from our experience with the Hoyle state that such a cluster state cannot be described within the shell model alone, see \cite{Lovas} \cite{astier}. In the infinite matter section, we have already considered a situation where a single $\alpha$ particle is imbedded in an uncorrelated Fermi gas. This was the case with respect to the critical temperature. For $^{212}$Po we can consider our treatment of the single $\alpha$ case even at zero temperature. Of course we cannot use the ansatz (2) as is, since it corresponds to a c.o.m. wave function which is a plane wave $e^{i{\bf K}{\bf R}}$. In position space (2) namely reads $\Psi_{1234} \propto  e^{i{\bf K}{\bf R}} 
\psi_{int.}(|{\bf r}_i - {\bf r}_j|)$ where $\psi_{int.}$ is the intrinsic wave function depending only on the relative coordinates. In a finite nucleus, instead of the plane wave, one would have to use a wave packet considering the following wave function

\begin{equation}
\Psi_{1234} \propto \Phi({\bf R})
\psi_{int.}(|{\bf r}_i - {\bf r}_j|)
\end{equation}

where  $\psi_{int.}$ and $\Phi$ are to be determined 
variationally upon inserting (12) into eq (1) where all indices and 
ingredients correspond now to the mean field of $^{208}$Pb. For instance 
the $\varepsilon_i$ correspond to the HF energies and the $f_i$ are to be 
replaced by the HF occupation numbers at $T=0$. In addition for the $\psi_{int.}$, one 
could use, as for the study of the Hoyle state, a gaussian wave function with 
the width parameter $b$. Then the single unknown would be the spherical wave 
function $\Phi(R)$ to be determined variationally where $R$ is the c.o.m. 
coordinate of the $\alpha$ with respect to the center of the Pb core. Our 
knowledge that $\alpha$'s only can exist at very low density incites us to believe that 
this $\Phi$ wave function should be peaked rather far out in the surface of 
the Pb nucleus. If true, this would be a nice explanation of a preformed 
$\alpha$ particle in the nuclear surface. Adding more than one $\alpha$ to 
the Pb core may suggest that there could exist some sort of $\alpha$ 
condensate in the surface in a fluctuating state. However, as we know, adding 
more $\alpha$'s to the Pb core leads to deformed nuclei and the treatment 
of $\alpha$'s in the surface becomes a much more delicate subject.

\section{ Conclusions} As we have seen, the existence of $\alpha$ gas and $\alpha$ condensate states 
in nuclear systems where the $\alpha$'s play practically the role of 
elementary bosons, is fascinating. Nuclear physics is at the forefront 
of this kind of physics. In the future, experiments with cold atoms 
trapping four (or more) different kinds of fermions may also open wide perspectives in 
the field of cluster physics. For more reading on $\alpha$ cluster states, we invite the reader to consult our review article \cite{book}.

\ack

This contribution is in collaboration with, Y. Funaki, H. Horiuchi, G. R\"opke, T. Sogo, A. Tohsaki, and T. Yamada.

\end{document}